\documentclass[12pt]{iopart}
\usepackage[pdftex]{graphicx}

\begin{document}

\title[Head-on collision of ion-acoustic solitary waves]{Head-on collision of ion-acoustic solitary waves in a collisionless plasma}

\author{Yu V Medvedev}

\address{Joint Institute for High Temperatures of Russian Academy of Sciences, Moscow, 125412, Russia}

\begin{abstract}
{The head-on collision of ion-acoustic solitary waves in a collisionless plasma with cold ions and Boltzmann electrons is studied. It is shown that solitary waves of sufficiently large amplitudes do not retain their identity after a collision. Their amplitudes decrease and their forms change.  Dependences of amplitudes of the potential and densities of ions and electrons after a head-on collision of identical solitary waves  on their initial amplitude are presented.}
\end{abstract}

\nosections

The formation and propagation of ion-acoustic solitary waves in plasmas have been extensively investigated over a period of more than fifty years. Of particular interest is the interaction (collision) of solitary waves with each other. The investigations were carried out in  two-component electron-ion plasmas 
[1--8] and in multicomponent plasmas that  contain,  in addition to positive ions and electrons, other components such as negative ions, positrons or dust particles 
[9-22]. A considerable number of papers were devoted to the study of plasmas with non-Maxwellian particle distributions 
[3, 6, 12, 16--20, 22]. Analytical 
[3, 6, 12--22] and numerical 
[2, 4, 5, 7, 8, 23] methods were used  to investigate solitary waves. Experiments with collisions of solitary waves were described in 
[1, 9--11].

In most papers, the studies of collisions of solitary waves were performed for the case of not very large amplitudes and were aimed at determining the main results of the collision: the phase shifts and the trajectories of solitary waves after the collision.
In this paper we are dealing with head-on collisions of solitary waves of large amplitudes and  the main attention is focused on the very  process of solitary waves collision. The collision  is accompanied by a change in the nature of the motion of  plasma particles over which the solitary wave propagates. 
It is shown that solitary waves of sufficiently large amplitudes do not retain their identity after a collision. Their amplitudes and propagation speeds decrease, and their forms change. In  the collision, some of the ions are accelerated to considerable velocities and leave the solitary wave regions. The range of amplitudes of solitary waves in which  they can be considered as solitons is found.
 Dependences of amplitudes of the potential and densities of ions and electrons after a head-on collision of identical solitary waves  on their initial amplitude are also presented.
 
The problem of the interaction of solitary waves is considered in the one-dimensional approximation and is formulated as follows. At the initial time $t=0$ in the region $[-L, L]$ there is a collisionless electron-ion plasma with singly charged cold ions and electrons obeying the Boltzmann distribution. The plasma has a uniform distribution of quantities in space everywhere, except for two small regions. In these regions there are two ion-acoustic solitary waves of equal amplitude: solitary wave 1  with its amplitude coordinate  $x=-a$ and propagation velocity $D>0$ and solitary wave 2 with its amplitude coordinate  $x=a$ and propagation velocity $-D$. So the solitary waves propagate towards each other. At some time $t>0$ they collide with each other in the neighborhood of the point $x=0$ and then diverge. 

To study this process we use numerical simulation by the particle-in-cell (PIC) method. The initial distributions of the quantities in the solitary waves are calculated using the technique described in \cite {Medvedev, Book}. These distributions are implemented in the simulation model by specifying the  coordinates and velocities of the model particles. At the boundaries $ x = \pm L $, the specular reflection condition is set for particles. The Poisson equation is solved with zero electric field at the boundaries. The electric potential $ \varphi (x, t) $ is measured from the zero value in the unperturbed region. In the simulations, the solitary waves formed at the initial time propagate without any changes until they collide.

Below all quantities are given in dimensionless form.  The values:
\[ (m_i/4\pi e^2n_{e0})^{1/2},\qquad (T_{e0}/4\pi e^2n_{e0})^{1/2}, 
\]
\[ (T_{e0}/m_i)^{1/2},
\qquad n_{e0}, \qquad T_{e0}/e.
\]
are used as units of time, length, speed, density, and potential, respectively. Here $e$ is the absolute value of the electron charge, $m_i$ is the mass of the ion, and $n_{e0}$ and $T_{e0}$ denote the electron density and temperature (in energy units) in the unperturbed region.

Consider the results of numerical simulations obtained for $L=100, \; a=60$.   Figure  1 shows the dependence of the  potential maximum $\varphi_m $ on time in the region $ x \le 0 $ during the propagation and the head-on collision of two identical solitary waves for two cases of the initial  amplitude $ \varphi_ {m0}$. 
 The dependence $\varphi_m(t)$ in the region $x>0$ have exactly the same form due to the symmetry of the problem.
 In the case of $ \varphi_ {m0} = 0.85 $, there are several narrow peaks on the background of the constant value $ \varphi_m = \varphi_ {m0} $. It is easy to understand that the time intervals during which $ \varphi_m $ remains unchanged correspond to the stationary propagation of the solitary wave with amplitude $ \varphi_{m0} $, and the first peak corresponds to the head-on collision of the solitary waves at the point $ x = 0 $. In this case after the collision the maximum of the potential is equal to the initial amplitude  of the potential $ \varphi_m = \varphi_ {m0} $.
 
\begin{figure}\centering
\includegraphics[viewport =  184 548 419 784,width=235pt]{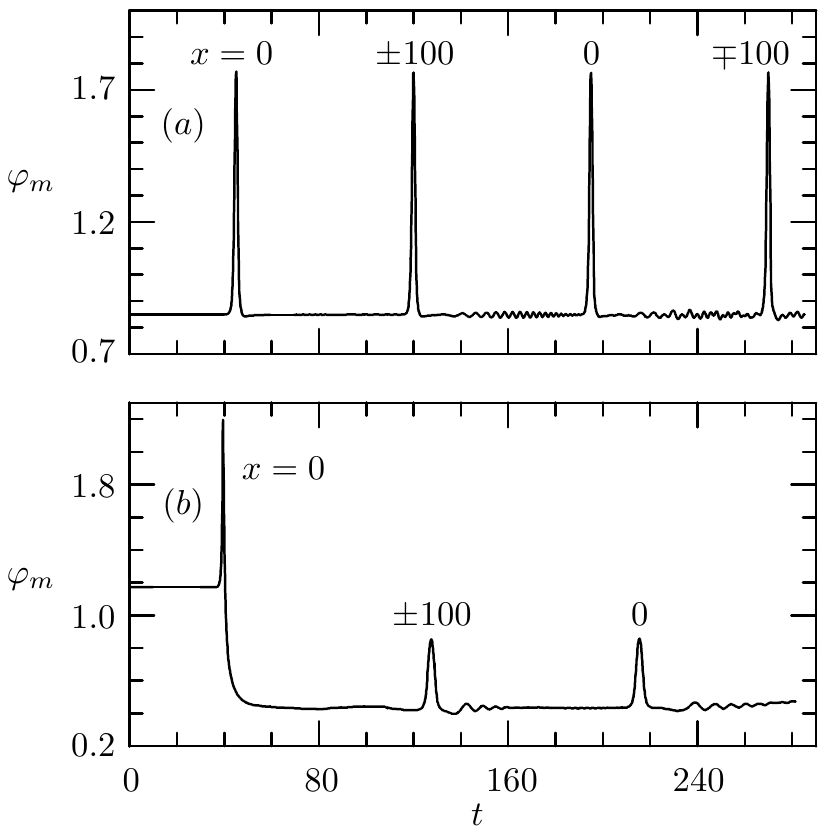}
\caption{The maximum of the potential  $\varphi_m$ versus time for two cases: (a) $\varphi_{m0}=0.85$ and (b) $\varphi_{m0}=1.175$. The indicated coordinates of the peak values of $\varphi_m$ correspond  to the collision of the solitary waves $(x=0)$, the reflection of the solitary waves from the  boundaries $(x=\pm 100)$, the second collision of the solitary waves $(x=0)$ and the second reflection of the solitary wave  from the boundaries $(x=\mp 100)$
} \label{fig1}
\end{figure}

Since the colliding solitary waves are exactly the same, the result of their head-on collision can be interpreted either as the passage of solitary waves through each other or as their reflection from each other. In the first interpretation, solitary wave 1  propagating in the region $ x <0 $ before the collision  propagates in the region $ x> 0 $ after the collision, and conversely,  solitary wave 2  propagating in the region $ x> 0 $ before the collision propagates in the region $ x < 0 $ after the collision. After the collision at the point $x=0$, solitary wave 1 (solitary wave 2) then collides with the boundary $ x = L $ ($x=-L$) that specularly reflects plasma particles.  Reflection of particles at the boundary leads to the fact that the entire solitary wave is reflected from the boundary.  The second peak in Figure 1a corresponds to these reflections.

Then the solitary waves again propagate  towards each other and collide a second time at the point $x=0$ (the third peak in Figure 1a). The fourth peak in Figure 1a corresponds to the next reflection of the solitary waves from the boundaries. As we see, in the case of $ \varphi_ {m0} = 0.85 $ the collision of solitary waves with each other or with the boundaries does not lead to a change in their amplitude and, accordingly, in their forms. Note that the preservation of the identity of solitary waves of small amplitude, when they are reflected from the boundary on which the particles  should be reflected, was observed in a numerical experiment also for  a chain of solitary waves \cite{PPCF}.

A completely different result is obtained in the case of  identical solitary waves with  $ \varphi_ {m0} = 1.175 $ (Figure 1b). Here, after the first head-on collision at the point $x=0$, the amplitudes of both solitary waves drop abruptly. The amplitudes are even noticeably smaller ($ \varphi_m \sim 0.42 $) than the solitary wave amplitudes in the case considered in Figure 1a. Obviously, the speed of propagation decreases and the form of each solitary wave changes, in particular, its width increases. Due to the larger width of the solitary waves formed after the first collision,  the subsequent collision of each solitary wave with the specularly reflecting boundary (the second peak in Figure 1b) takes a longer time interval than in the case shown in Figure 1a. It can be seen that subsequent collisions of the  solitary waves with boundaries or with each other at the point $ x = 0 $ (the second and the third peak in Figure 1b) do not change their amplitudes because these amplitudes are small enough. Essential changes occur only at sufficiently large  amplitudes.
 
Obviously,  a head-on collision of identical solitary waves can be regarded as a reflection of the solitary waves from each other. In this interpretation, after the head-on collision, each solitary wave changes the propagation direction, but remains in the original region $x<0$ (solitary wave 1) or in region $x>0$ (solitary wave 2).    
Let us use this idea for a more detailed consideration of the head-on collision of two solitary waves with the same amplitudes.  Figure 2 shows the phase planes of the ions $ (x, v_ {i1}) $ in  solitary wave 1 near the point $ x = 0 $ for those instants of time  that are close to the time corresponding to the first peak in Figure 1. If the phase planes of the ions $ (x, v_ {i2}) $ in solitary wave 2 were shown in Figure 2 for the same instants, they would be symmetric about the point $ x = 0, v_ {i1} = 0 $.

\begin{figure}[b]\centering
\includegraphics[viewport =  192 548 419 785,width=227pt]{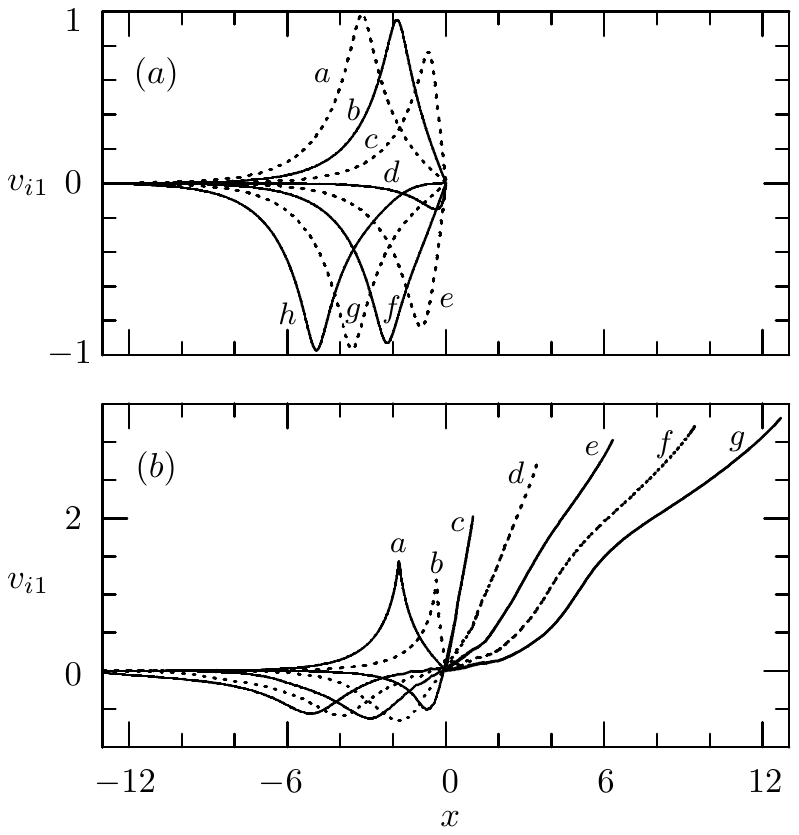}
\caption{The phase planes of ions in  solitary wave 1 for two cases: (a) $\varphi_{m0}=0.85$, the curves $a, b,\ldots, h$ correspond to the times $t=42, 43, \ldots, 49$; (b) $\varphi_{m0}=1.175$, the curves $a, b,\ldots, g$ correspond to the times $t=38, 39, \ldots, 44$
} \label{fig2}
\end{figure}

Figure 2a shows that in the case of solitary waves of not very large amplitude $ \varphi_ {m0} = 0.85 $, the front side of the ion velocity profile becomes steeper with time (curves $ a, b, c $), and then the ion velocities change sign (curves $ d, e $). Gradually the maximum ion velocity in absolute value reaches the same absolute value as before the collision (curves $ f, g, h $). With a change in the directions of ion velocities, the solitary waves change their directions of propagation: in the region $ x <0 $  reflected solitary wave 1 propagates to the left, and in the region $ x> 0 $  reflected solitary wave 2 propagates to the right. In the case of  $ \varphi_ {m0} = 0.85 $, the velocities of all plasma ions involved in the collision of the solitary waves return to the initial zero value after the passage of the solitary waves. 

A different nature of the ion motion can be seen at the larger initial solitary wave amplitude $ \varphi_ {m0} = 1.175 $ (Figure 2b). In this case the front side of the ion velocity profile in solitary wave 1 first steepens (curves $ a $ and $ b $), and then the ion velocities change sign (curve $ c $). The distribution of the negative ion velocities acquires a form similar to a solitary wave (curves $ c $ - $ g $ in the region of negative velocities). However, some ions do not change sign of velocity. These ions are in the region where the electric field increases with time as the solitary waves converge, and are appreciably accelerated. They go to the region $x>0$ (curves $ c $ - $ g $ in the region of positive velocities).
The ion velocity profile in solitary wave 2 propagating along the region $ x> 0 $ behaves in a similar manner. During the collision of the solitary waves,  most ions in this region change their direction of motion to the opposite direction, but some of the ions having  sufficiently high velocities continue to move in the same direction. Moreover, their velocities increase and they go to the region $x<0$. 

\begin{figure}[b]\centering
\includegraphics[viewport = 185 548 419 784,width=234pt]{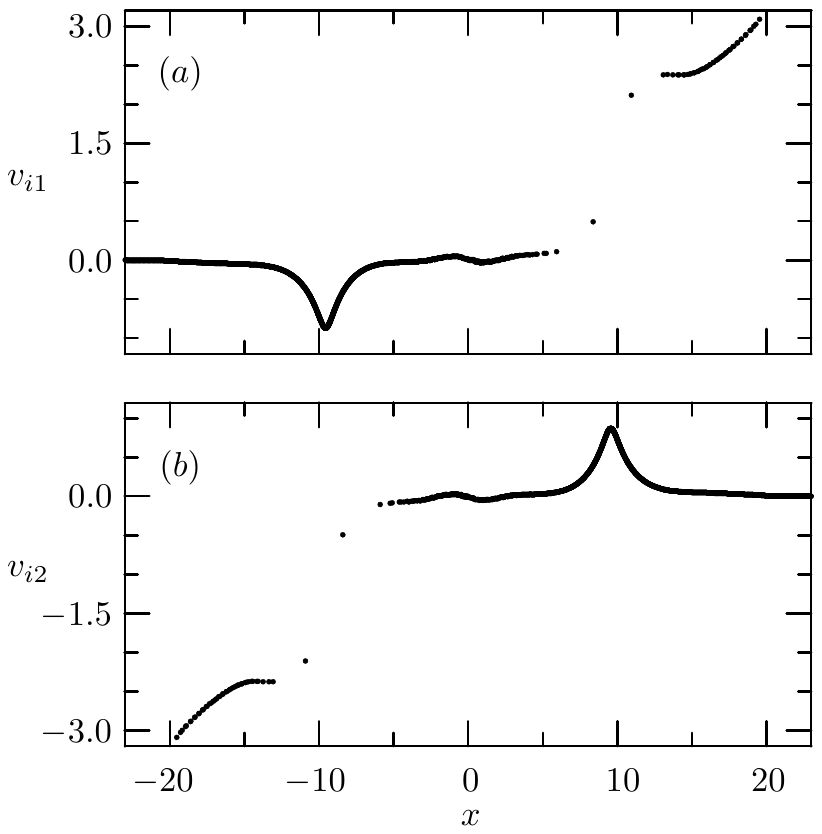}
\caption{The phase planes of ions after the collision at $t=50$ in the case of $\varphi_{m0}=1.0$. (a) ions in  solitary wave 1, (b) ions in  solitary wave 2
} \label{fig3}
\end{figure}

In this case, a peculiar exchange of ions takes place: the accelerated ions initially located in the region  $ x <0 $ go to the region $ x> 0 $, and  the accelerated ions initially located in the region  $ x >0 $ go to the region $ x< 0 $. The accelerated ions have enough energy to leave the region of the solitary wave and then they move freely through the unperturbed plasma.  Figure 3 illustrates the separation of a bunch of accelerated ions from the bulk of their  `` own '' ions, along which the reflected solitary wave of reduced amplitude propagates in the opposite direction. The velocities of the accelerated ions lie in the interval from 2.4 to 3.0. It can be seen that the accelerated ions overtake the `` strange '' solitary wave propagating in the same direction. These ions take energy from the solitary waves. As a result the amplitudes  of the solitary waves decrease and the spatial distributions of the quantities in each solitary wave change.

\begin{figure}\centering
\includegraphics[viewport = 184 619 419 784,width=235pt]{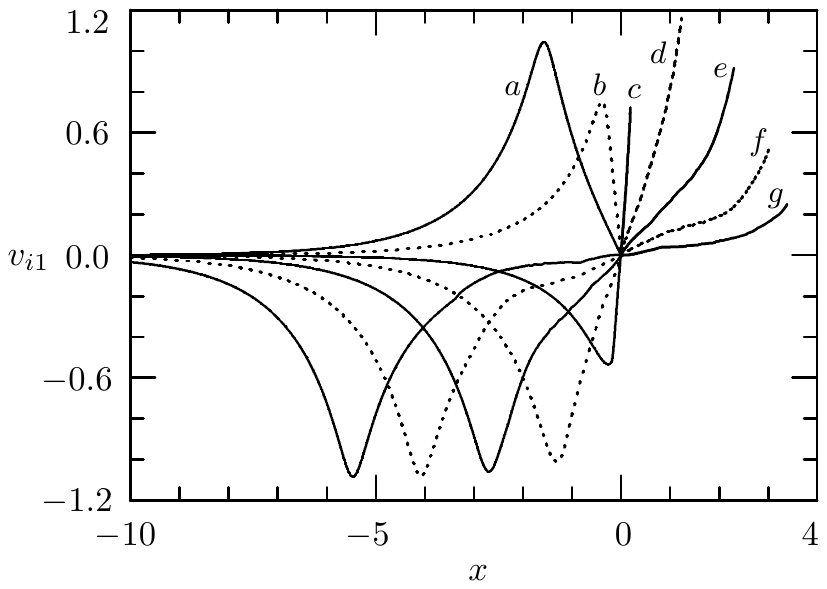}
\caption{The phase planes of ions in  solitary wave 1 in the case of $\varphi_{m0}=0.925$. The curves $a, b,\ldots,g$ correspond to the times $t=42, 43, \ldots, 48$
} \label{fig4}
\end{figure}

From numerical experiments performed for different amplitudes of the solitary waves $ \varphi_ {m0} $ it is established that for $ \varphi_ {m0} \le 0.925 $ the solitary waves formed after the head-on collision practically do not differ from the solitary waves before the collision. In the limiting case of $ \varphi_ {m0} = 0.925 $, in contrast to the case of $ \varphi_ {m0} = 0.85 $ (Figure 2a),  the formation of a bunch of accelerated ions is  observed, as well as the exchange of ions between regions $ x <0 $ and $ x> 0 $ (Figure 4). However, soon the accelerated ions are decelerated by the electric field of the solitary wave (curves $ e, f, g $ in Figure 4). The solitary wave passes through these ions, leaving them immovable. The amplitudes of the solitary waves remain unchanged after this collision.

At higher  initial amplitudes of the solitary waves $ \varphi_ {m0} $, their amplitudes after collisions are not conserved. Figure 5 shows the amplitudes  of the potential $ \varphi_m $ and densities of ions $ n_ {im} $ and electrons $ n_ {em} $  after a collision of solitary waves as  functions of $ \varphi_ {m0} $. It is seen that the larger the amplitudes of solitary waves before their collision, the smaller the amplitudes are after the collision. The figure confirms the presence of a noticeable space charge in the ion-acoustic solitary wave, which was discussed in \cite {Medvedev}. At $ \varphi_ {m0}> 0.925 $ the value of $ n_ {im} -n_ {em} $ decreases with increasing $ \varphi_{m0} $.
    
\begin{figure}\centering
\includegraphics[viewport =  192 617 439 784,width=247pt]{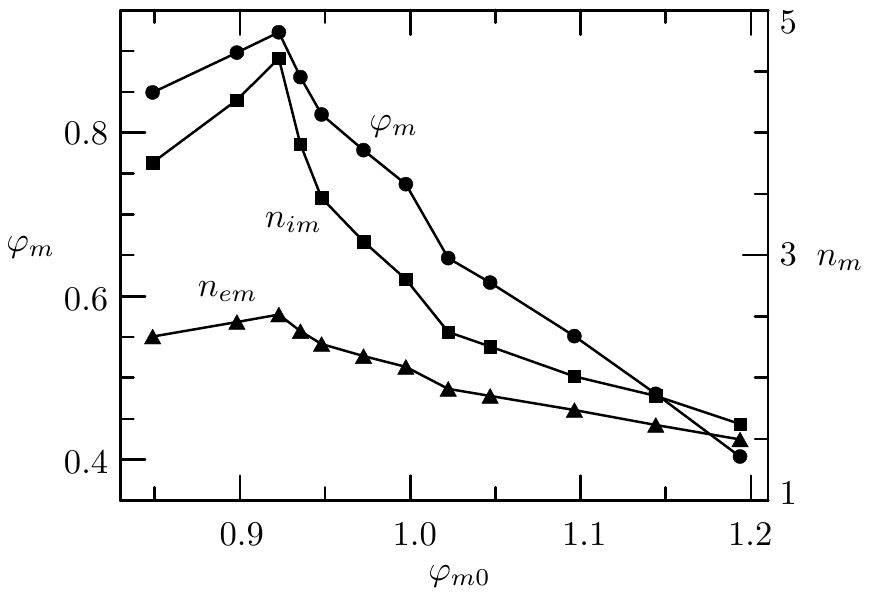}
\caption{The  amplitudes of the potential $\varphi_m$ and densities of ions  $n_{im}$ and electrons $n_{em}$ (right y-axis) after a collision of solitary waves versus  their initial amplitude  $\varphi_{m0}$
} \label{fig5}
\end{figure}

Thus, ion-acoustic solitary waves of sufficiently large amplitudes ($ \varphi_ {m0}> 0.925 $) do not survive after their  head-on collisions. A solitary wave that survives in a collision with another solitary wave is called a soliton \cite{Zabusky}. Beginning with Sagdeev's work \cite{Sagdeev}, it is known that an ion-acoustic soliton in a plasma with cold ions and Boltzmann electrons can exist at potential amplitudes up to the critical value $\varphi_{cr}\approx 1.256$. The results presented here show that the limiting  amplitude of the ion-acoustic soliton does not exceed $\varphi_{lm}\approx 0.925$. At higher amplitudes, a solitary wave can propagate in a plasma, but it is not a soliton, since its amplitude and form are not conserved in collisions.

\end{document}